\documentclass[pra,aps,preprint,showpacs]{revtex4}

\usepackage{mathrsfs}
\usepackage{amsfonts}
\usepackage{amsmath}
\usepackage{amssymb}
\usepackage{graphicx}
\usepackage{color}
\newcommand{\ket}[1]{|#1\rangle}

\begin{document}
\title{Oscillator Model of Spin}
\author{Yitian Ding\footnote{Email: dyt@mail.sdu.edu.cn}, Miaomiao Xu}
\affiliation{Department of Physics, Shandong University, Jinan 250100, China }
\date{\today}
\begin{abstract}

The Schwinger's representation of angular momentum(AM) relates two important
fundamental models, that of AM and that of harmonic oscillator(HO). However, the representation
offers only the relations of operators but not states. Here, by developing a graphic method
to calculate the states, we show that
the Schwinger's model is valid only with certain spin. With
the relation of states, we also demonstrate how the superposition states in AM
map to entangled states in HO. This study has promising applications in quantum computation,
and may cast light on the relation between superposition and entanglement.

\end{abstract}
\maketitle
\date{\today}

\section{Introduction}

Schwinger's oscillator representation of AM \cite{schwinger}
reveals connection between two important and fundamental mathematical models
in physics - AM and HO. The three operators constructed from
two pairs of boson creation and annihilation operators, which belongs to
two uncoupled HOs, satisfies the algebra of AM. It has been carefully investigated \cite{jpa1,jpa2} and broadly used 
in the study of fundamental particles \cite{particle}, condensed matter \cite{cmp}, atoms \cite{atom} and optics \cite{optics}.
AM-like algebra occurs mainly in the property of single particle, while HO-like that of field
theory, thus the representation also relates the theories of particles and fields.
As is well known, control of single particle's spin is useful but challenging \cite{farhi,petta,AM}, and
HO-like systems \cite{HO} may provide to simulate the process and to certain extent replace it.

Superposition of states has been the core contents to quantum mechanics \cite{dirac,zeilinger}. Most new phenomena
unique of quantum physics come from the uncertainty resulted from the superposition, for one physical quantity contains
the possibility of two values and result of measurement is totally unpredictable. Meanwhile, entanglement \cite{epr,bell,pan}
is one of the strangest behaviors in quantum physics, for it violates the law set in
relativity and contains superluminal property. The relation between these two important
concepts may reveal the essential of quantum physics.

In this Letter, we realize the oscillator model in finite AM number case, which we call
the spin model case. It requires the eigenvectors of AM operator in $x$ and $z$ direction. We
develop a graph, with careful definition of each element, to demonstrate the
process of solving these eigenvectors. Then, the realizable spin number and state relation are obtained.
Then we study possible modification to this state relation, and after it is proved
to be the only possible relation, we show that the AM's superposition states
is one case of HO's entangled states.

\section{HO, AM, and Schwinger's Representation}

First, we review certain facts of HO and AM, and set up the symbols used through the paper. In
the occupation number representation,
\[\begin{array}{l}
{a^\dag }\left| n \right\rangle  = \sqrt {n + 1} \left| {n + 1} \right\rangle \\
a\left| n \right\rangle  = \sqrt n \left| {n - 1} \right\rangle
\end{array}\]
in which the $a^\dag$ and $a$ are generation and annihilation operators
of particles, and $\left| n \right\rangle$ the $n$th eigenstate of number
operator $N=a^\dag a$ with eigenvalue $n$. The oscillator here is bosonic,
thus the commutation relation(CR) $\left[ {a,{a^\dag }} \right] = 1$ holds.
Meanwhile, the CR of AM reads $\left[ {{J_i},{J_j}} \right] = \hbar {\varepsilon ^{ijk}}{J^k}$,
which contains all the information of AM needed through this paper.

Then, the relation between these models is given. The Schwinger's representation of AM,
\begin{equation}\begin{array}{l}
{J_z} = {\textstyle{\hbar  \over 2}}({a^\dag }a - {b^\dag }b)\\
{J_x} = {\textstyle{\hbar  \over 2}}({a^\dag }b + {b^\dag }a)\\
{J_y} = {\textstyle{\hbar  \over {2i}}}({a^\dag }b - {b^\dag }a)
\end{array}\end{equation}
satisfies the algebra of AM operators, thus the connection of AM and HO
ensures the correspondence of two sets of operators. However, as we would see
later in this paper,
calculation on the magnitude of eigenstates, especially that of the
spin-$x$ one, shows the relation is physically realizable only with
certain spin numbers.

\section{Representation of Spin AM}

We use $\left| {m,n} \right\rangle  = \left| m \right\rangle  \otimes \left| n \right\rangle$
for the eigenstates of two uncoupled HOs, while $\left| m \right\rangle$ and
$\left| m \right\rangle$ are their own ones. Subsequently changed the operators,
$a \to a \otimes I,b \to I \otimes b$.
The spin-$z$ eigenstates are ${\textstyle{\hbar\over 2}}\left({m-n}\right)\left| {m,n} \right\rangle$,
when $m-n=1,-1$, correspondingly spin up and down. Then we calculate the spin-x eigenstates, in which we still
requires ${J_x}\left| x  \right\rangle  =  \pm {\textstyle{\hbar  \over 2}}\left| x  \right\rangle$.
After substituting $\left| x  \right\rangle$ with the superposition of
spin-$z$ eigenstates, $\sum\limits_{k,l \ge 0} {{C_{kl}}} \left| {k,l} \right\rangle$,
we obtain $({a^\dag }b + {b^\dag }a)\sum\limits_{k,l \ge 0} {{C_{kl}}} \left| {k,l} \right\rangle  =  \pm \sum\limits_{k,l \ge 0} {{C_{kl}}} \left| {k,l} \right\rangle$, which equals to,
\begin{equation}\begin{array}{l}
~~~~~\sum\limits_{k,l \ge 1}C_{kl}\left( \sqrt l \sqrt {k + 1} \left| {k + 1,l - 1} \right\rangle\right)\\
+~~\sum\limits_{k = 0,l \ne 0} {\sqrt l } {C_{0l}}\left| {1,l - 1} \right\rangle  + \left( {k \leftrightarrow l} \right)\\
=\pm \sum\limits_{k,l \ge 0} {{C_{kl}}} \left| {k,l} \right\rangle
\end{array}\label{eqofinterest}\end{equation}
To demonstrate what this equation mean, we develop a graph in which the
function of operator ${a^\dag }b + {b^\dag }a$ is shown.

\section{The graphic method and realizable spin numbers}

We denote the eigenstates of Spin-$z$ with a graph of squares, with each square
one state. The initial states of interest, the eigenstates of spin-$x$ composed of
that of spin-$z$, equal to one $n \times n$ square area. The final states, those
on the left-hand side of \ref{eqofinterest}, are two areas, each containing
$n^{2}$ squares, and in the upper right and lower left corner of the initial
area. The equation of eigenstate requires a selection of squares inside the initial
area as the initial state, such that the corresponding final state, obtained from the selected initial
one through operation of ${a^\dag }b + {b^\dag }a$, is identical to the initial one.

\begin{figure}
\begin{center}
\scalebox{0.4}{\includegraphics{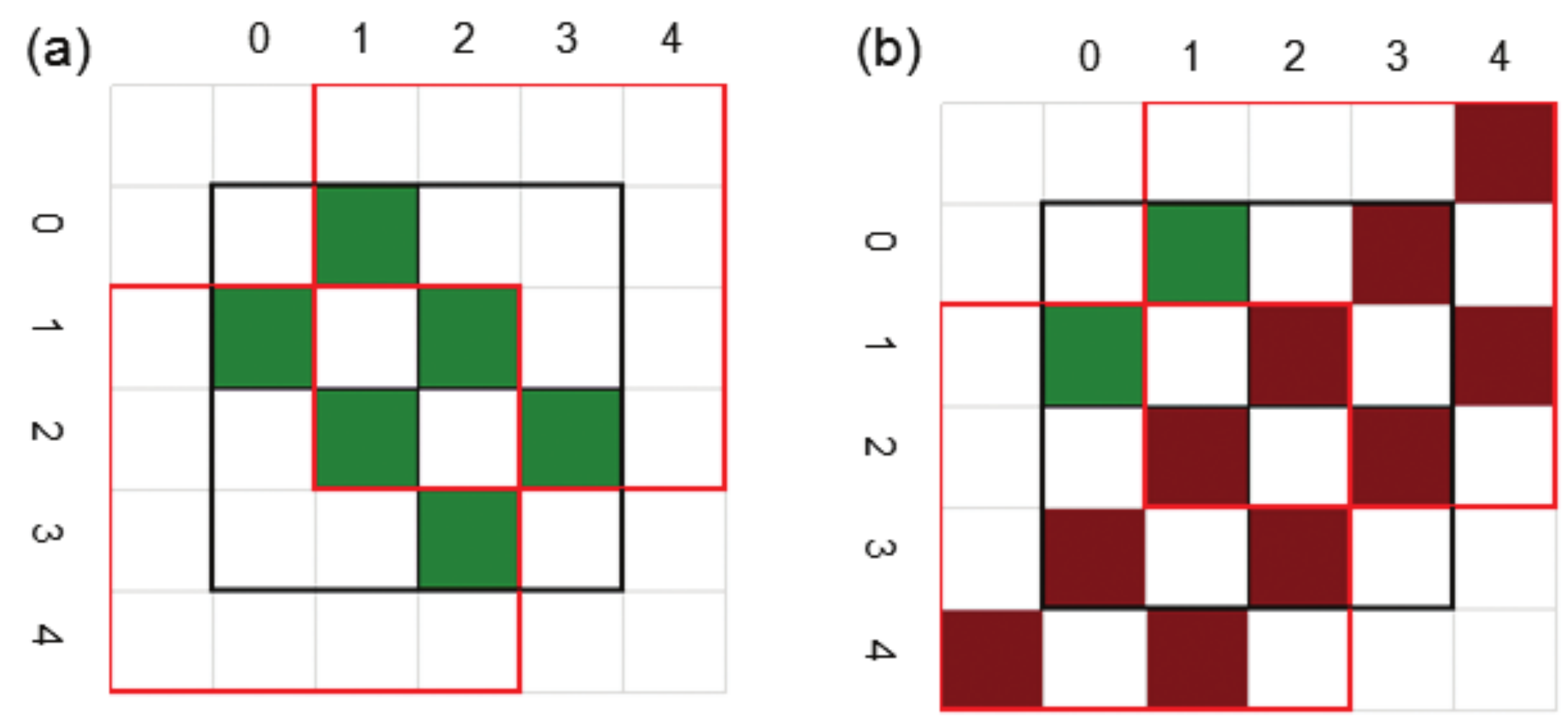}}
\caption{Graphic method on $4 \times 4$ initial area. The initial area is surrounded with black line and the corresponding final area with red line. The green squares in (a) are spin-z eigenstates, with bands (1,0)\&(0,1), (2,1)\&(1,2) and (3,2)\&(2,3). Aside from the upper left band in (b), others do not exist. (1,2) would result in (0,3), with the same magnitude, which could only be made zero with non-existed (-1,4); (2,3) would result in (1,4), which is not inside the initial area}
\end{center}
\end{figure}

The initial area shrink to a small selectable area, since only the eigenstates
of spin-$z$ with eigenvalues of $1$ or $-1$ could
serve as components of the initial state. The area of selectable squares shows
band-like structure. The bands which leaves blank squares inside the initial area
at each end would result in new states after the operation, and whose magnitude
could only be made zero if the magnitude of the corresponding initial ones are
zero. Meanwhile, the one at the lower right corner would result in states which
are outside of the initial area. Thus, the remaining selectable states are just the first
band in the upper left.

The calculation for spin $\frac 12$ particle of magnitude of
$\left| z \right\rangle$'s in $\left| x \right\rangle$
shows the up and down states for spin-$x$ is
$\left|x,\pm\right\rangle=1/\sqrt{2}\left(\left|z,+\right\rangle\pm\left|z,-\right\rangle\right)$.
However, calculation for the spin $1$ and $\frac 32$ shows the equivalent magnitude
zero. Study of the first band reveal the cause. The magnitude of states, from the
lower left to the upper left would be $C_{n,0},C_{n-1,1}...C_{1,n-1},C_{0,n}$. Through
operation of ${a^\dag }b + {b^\dag }a$, it should keep unchanged due to its eigenvector
characteristic. With the graph, the process of operation is equivalent to
$C_{n,0}=C_{n-1,1},C_{n-1,1}=C_{n,0}+C_{n-2,2},...$ and also
$C_{0,n}=C_{1,n-1},C_{1,n-1}=C_{0,n}+C_{2,n-2},...$.
The magnitude of states then reduce to
\[\begin{array}{l}C_{n,0},C_{n,0},0,-C_{n,0},-C_{n,0},0,C_{n,0},...\\
...,C_{0,n},0,-C_{0,n},-C_{0,n},0,C_{0,n},C_{0,n}\end{array}\]
with both a period of $6$. The only situation in which the magnitude are not all zero is when $0$ from
the two sides coincide, which gives
\begin{equation}2s+1=3n+2,n=0,1,2,...\end{equation}
in which spin with half-integer or integer equals to $n$ an even or odd number.

This is a test of whether the spin-$\frac{1}{2}$ state in spin-$x$ representation could be realized
in a system with spin-$\frac{n}{2},n=1,2,3,...$. Since the particle with an integer spin would not
acquire it, we now come to study the spin-$1$ case. A similar series $C_{n,0},2C_{n,0},3C_{n,0},4C_{n,0},...$
shows that the magnitude should all be $0$. Thus an integer spin number could not be realized. If we
study further, an identical phenomenon would make spin $\frac{3}{2}$ impossible. Actually, the only
realizable spin-$x$ case is the spin-half state for
\begin{equation}s=\frac{6n+1}{2},n=0,1,2,...\end{equation}
and a full representation of quantum system, or in other words, a complete modeling would
only occur for spin-half particles like nucleons \cite{particle}.

\section{uniqueness of commutation relation}

The ladder operators of AM satisfy $[J_{+},J_{-}]=2\hbar J_{z}$
and $[J_{z},J_{\pm}]=\pm \hbar J_{\pm}$. In the following, we set $\hbar=1$. The
new commutator assumed is
\begin{equation}[b,a]=c,[b,a^\dag]=d\end{equation}
depicting the interaction of two HOs, and the situation of interest keep the algebra of AM unchanged.
The method is iterative: the commutation relation of AM is maintained; and with equation
$[J_{+},J_{-}]=2\hbar J_{z}$ and new commutation of the HO operators,
a new $J_{z}$ -- $J_{z}'$ -- is calculated. Then with the second one
$[J_{z},J_{\pm}]=\pm \hbar J_{\pm}$ and the new $J_{z}'$ obtained us $J_{+}'$. By means of iteration, this $J_{+}'$ is substituted
into the first equation again for a third $J_{z}$ -- $J_{z}''$. The process goes on till no extra term
occurs in either new $J_{z}$ or new $J_{+}$.

In the first step of iteration,
\begin{equation}\begin{array}{l}J_{z}'=\left[J_{+},J_{-}\right]=a^\dag a-b^\dag b+ca^\dag b^\dag+c^\star ab\\
J_{+}'=[J_{z}',J_{+}]=(1-|c|^2)a^\dag b-(ca^\dag a^\dag-c^\star bb)\\
~~~~~~~~~~~~~~~~~~~~~~-\frac{d}{2}(a^\dag a + b^\dag b+ca^\dag b^\dag + c^\star ab)\end{array}\end{equation}
the coefficient of $a^\dag b$ only concerns $c$, and since no more $a^\dag b$
is allowed due to the iteration nature, $c=0$. $J_{z}'=J_{z}$ and $J_{+}'$ simplifies to
$a^\dag b-\frac{d}{2}(a^\dag a + b^\dag b)$. Since $J_{z}$ does not change, we iterate just with
$[J_{z},J_{\pm}]=\pm \hbar J_{\pm}$. In the second step,
\begin{equation}J_{+}''=[J_{z},J_{+}']=(1-\frac{|d|^2}{2})a^\dag b+\frac{d^2}{2}b^\dag a\end{equation}
thus $d=0$ for the same reason, which means the two sets of operators should commute with each other.
We then come to the conclusion that with coupling through commutation, the AM model cannot be built.

\section{coupling of two HOs with entanglement}

We now show that the generally accepted uncoupled oscillators are actually coupled with each other
with entanglement. For a random state in spin-$z$ representation of spin-half particle $\ket{\psi_{z}}=A\ket{z+}+B\ket{z-}$, which is
a superposition of two eigenstates, one would have in the oscillator model
\begin{equation}\ket{\psi_{z}}=A\ket{1,0}+B\ket{0,1}\end{equation}
which is actually an entangled state between the two oscillators. Similarly for spin-$x$, one would have
in $\ket{\psi_{x}}=C\ket{x+}+D\ket{x-}$ also entanglement
\begin{equation}\ket{\psi_{x}}=\frac{1}{\sqrt{2}}[(C+D)\ket{1,0}+(C-D)\ket{0,1}]\end{equation}
except for special cases when $C=\pm D$. Under the model offered by Schwinger, the superposition states in AM could be represented by the entangled states in HO. Since we have
proved previously that no other algebra in bosonic HOs is able to support the AM math structure,
the statement that the superposition of AM states is equivalent to one kind of entanglement in HO generally applies.

\section{discussion and conclusion}

From FIG.1 we could see, the AM model constructed from HOs applies with
both finite and infinite AM number, for the only part of significance is the first band.
The infinite situation is just to extend the graph to lower-right, which contains no new states
available. Thus it is meaningless to generally talk about operator relation without state relation
when it comes to certain spin number.

Moreover, the two HOs are completely not uncoupled. Instead, they are entangled. We have also
proved that the entanglement is inevitable, for the commutation relation is proved unique. The
superposition in AM is at least one kind of entanglement in HO, and whether they are equivalent
need further discussion.

In summary, we studied the state relation in the HO model of AM. By using a method based on graph, we visualized the realization process
from HO states to AM states, and obtained the realizable spin numbers and state relation. Besides, we have proved the superposition
states of AM is equal to one case of entangled states in HO, and the two HOs are unavoidably entangled.
This theoretical work may assist the understanding of common particles with spin $\frac{1}{2}$. It is also potentially useful
to quantum computation, since the broadly used electrons could be modeled with bosonic
systems. For the connection between AM's superposed states and HO's entangled states, it
needs further study whether in general systems there is a relation.

\end{document}